\documentclass[aps,showpacs,nofootinbib]{revtex4}

\usepackage{epsfig}
\usepackage{epsf}
\usepackage{bm}                 
\usepackage{amsmath}
\usepackage{amssymb}
\usepackage{dcolumn}
\usepackage{graphicx}
\usepackage{xcolor}
\usepackage[small,bf, format=plain, width=.75\textwidth, justification=RaggedRight]{caption}[2008/04/01]
\input amssym.def
\input amssym

\font\bbold=msbm10

\newcommand{\BR}{\mbox{\bbold R}}

\begin{document}

\title{On the Gravitational Inverse Problem}

\author{Ulvi Yurtsever}\thanks{Corresponding Author:
ulvi@phys.lsu.edu}
\affiliation{MathSense Analytics, 1273 Sunny Oaks Circle, Altadena, CA 91001 and
\\ Hearne Institute for Theoretical Physics, Louisiana State University,
Baton Rouge, LA 70803}
\author{Caren Marzban}
\affiliation{Applied Physics Laboratory, University of Washington, Seattle,
WA 98195 and
\\ Department of Statistics, University of Washington, Seattle, WA 98195}
\author{Marina Meil\v{a}}
\affiliation{Department of Statistics, University of Washington, Seattle, WA 98195}

\date{\today}

\begin{abstract}
We discuss some mathematical aspects of the problem of inverting
gravitational field data to extract the underlying mass distribution.
While the forward problem of computing the gravity field from a given
mass distribution is mathematically straightforward, the inverse
of this forward map has some interesting features that make
inversion a difficult problem. In particular, the forward map
has an infinite-dimensional kernel which makes the inversion fundamentally
non-unique. We characterize completely
the kernels of two gravitational forward maps, one mapping
mass density to the Newtonian
scalar potential, and the other mapping mass density to
the gravity gradient tensor, which is the quantity most commonly
measured in field observations. In addition, we present some
results on unique inversion under constrained conditions, and
comment on the roles the kernel of the forward map and
non-uniqueness play in discretized approaches
to the continuum inverse problem.
\end{abstract}


\maketitle

{\bf \noindent Weighing the shape of a gravitating body}

More than thirty years ago, Mark Kac asked
``Can you hear the shape of a drum?" meaning: do two distinct planar domains
always have distinct spectra of eigenvalues
for their respective Laplace operators (acting on functions) with
the usual Dirichlet (or Neumann) boundary conditions? If the answer is yes,
the shape of a ``drum" can be inferred by hearing its spectrum (characteristic
sound), if the answer is no, then two distinctly shaped
drums may have identical spectra (in which case they
are called ``isospectral domains") \cite{Kac1966,Conway1994}.

Kac's article~\cite{Kac1966}
stimulated a long line of research which eventually
settled his question in the negative: There do exist isospectral domains (and,
more generally, isospectral Riemann surfaces and isospectral Riemannian manifolds in
higher dimensions) which are {\em not} isometric. In other words, the spectral
inverse problem is ill-defined, subject to a fundamental ambiguity which
can be precisely characterized~\cite{berger}.

A similar ambiguity plagues the gravitational inverse problem,
that is, the problem of inferring the precise shape of a mass
distribution by observing its distant gravitational field.

The gravitational inverse problem is the problem of inverting the gravitational
forward map, which we take to be a map sending a compact supported
mass distribution to a gravity observable: in practice, the observable could be either
the Newtonian gravitational potential or gravity gradients. 

More precisely, and focusing on the gravity potential $\Phi$
for the moment, what we will mean
by the gravitational inverse problem is the following: Given a
spherical region $B_R=\{\vec{r} : |\vec{r}| < R \}$ of radius $R$ in ${\BR}^3$,
and a solution $\Phi(\vec{r})$ (the gravitational potential in free space)
of the Laplace equation ${\nabla}^2 \Phi=0$ outside the region $B_R$ (i.e.\
for $|\vec{r}|>R$) which vanishes at infinity, find a
mass density distribution $\rho(\vec{r}\, ')$ supported
inside $B_R$ which gives
rise to $\Phi(\vec{r})$ in the exterior region outside $B_R$. In plainer
language, find a $\rho(\vec{r}\, ')$ with support inside $B_R$ such that
\begin{equation}
\Phi(\vec{r})= - \, G \int_{B_R} \frac{\rho(\vec{r}\, ')}{|\vec{r} - \vec{r}\, '|}
\, d^3 r' \; \; \; \; \; \; \; \; \; {\rm for} \; r > R \; .
\end{equation}

~

{\bf \noindent Kernel of the forward map onto the gravitational potential}

Equation (1) of course represents the unique solution to the ``forward
problem" where one searches for a solution $\Phi$ to
${\nabla}^2 \Phi = 4 \pi G \rho$ with vanishing boundary
conditions at infinity. Of key interest is the ``kernel" of this
(linear) forward map, i.e.\ the set of mass distributions $\rho$ supported inside
$B_R$ that are mapped to a potential $\Phi$ via Eq.\,(1) which identically
vanishes outside $B_R$.

{\noindent \bf \underline{Theorem 1}}: The kernel of the forward map Eq.\,(1)
mapping mass distributions $\rho$ supported in $B_R$ to solutions
of Laplace equation outside the region $B_R$ (i.e.\
for $|\vec{r}|>R$)
is precisely functions $\rho$ satisfying
\begin{equation}
\rho = {\nabla}^2 \chi \; ,
\end{equation}
where $\chi(\vec{r})$ is any (sufficiently
smooth) function on $\BR^3$ with support inside $B_R$ (i.e.\
$\chi(\vec{r})=0$ for $r>R$).
In other words, if $\rho$ is a solution of the inverse problem
for a given exterior potential $\Phi$, then $\rho + \nabla^2 \chi$ is
also a solution for any $\chi \in {C_0}^{\! \alpha}(B_R)$,
where $\alpha$ is a sufficiently large integer. Normally,
$\alpha \geq 2$ should be sufficient, but smoothness is not
a key issue; in particular, $\chi$ can even
be a distribution if point-mass (delta-function)
singularities need to be allowed in the problem.

\noindent {\bf Proof} in one direction is easy: Every function in the
kernel is given by the forward image of a function of the kind Eq.\,(2).
To prove this, let $\Phi$ be a function
belonging to the kernel, i.e.\ let $\Phi$ vanish outside $B_R$. Put
\[
\chi \equiv \frac{1}{4 \pi G} \Phi \; .
\]
Then $\chi \in {C_0}^{\! \alpha}(B_R)$ and $\rho \equiv \nabla^2
\chi$ satisfies the Laplace equation ${\nabla}^2 \Phi = 4 \pi G \rho$
everywhere (with vanishing boundary
conditions at infinity). Therefore, $\Phi$ satisfies Eq.\,(1) with this $\rho$,
which is what we needed to prove.

Conversely, let $\rho$ be a density
distribution supported inside $B_R$ such that $\rho = \nabla^2 \chi$
for some $\chi \in {C_0}^{\! \alpha}(B_R)$. Then, according
to Eq.\,(1), the gravitational
potential $\Phi$ which the forward map sends $\rho$ onto satisfies
\begin{equation}
\Phi (\vec{r}) =
- \, G \int_{B_R} \frac{\nabla^2 \chi (\vec{r}\, ' )}{|\vec{r} - \vec{r}\, '|}
\, d^3 r' \; \; \; \; \; \; \; \; \; {\rm for} \; r > R \; .
\end{equation}
To show that the right hand side of Eq.\,(3) is in the kernel of the forward map,
i.e., that it vanishes for $r>R$, use
Green's identity:
\begin{equation}
\int_{B} (U \nabla^2 V - V \nabla^2 U) \, d^3 r
=\int_{\partial B} \left( U \frac{\partial V}{\partial n}
- V  \frac{\partial U}{\partial n} \right) d \sigma \; ,
\end{equation}
where $B$ is any region bounded by the surface $\partial B$, and $U, \;
V$ are arbitrary functions on $\BR^3$. Applying Eq.\,(4) with $B$ taken
as the region $B_R$, $U(\vec{r}\, ') \equiv
1/|\vec{r}-\vec{r}\, '|$, and $V (\vec{r}\, ' ) \equiv \chi (\vec{r}\, ' )$,
and noting
that $\nabla^2 (1/|\vec{r}-\vec{r}\, '|) =0 $ when $r>R$ and $r' < R$,
it immediately follows
that the right hand side of Eq.\,(3) vanishes outside $B_R$ (i.e.\ for
$r>R$). This completes the proof of Theorem 1.

~

{\bf \noindent A geometric interpretation of the kernel:}

One way to conceptualize the kernel of the gravitational
(potential) forward map
is to note that the geometric
freedom of choice in the inverse ``datum" $\Phi(\vec{r})$ is that of choosing an
arbitrary function on the two-sphere $S^2$:

{\noindent \bf \underline{Theorem 2}}: Let $\Phi(\vec{r})$
be any solution of the free-space Laplace equation
${\nabla}^2 \Phi=0$ outside the region $B_R$
which vanishes at infinity, as in the formulation of
the gravitational inverse problem. Then $\Phi$ is completely determined
outside $B_R$ by its values on any two-sphere $S_{R_1}$ of radius
$R_1 >R$ (or, more generally, by its values on any closed surface which encloses
$B_R$).

\noindent {\bf Proof:} This is really a restatement of a standard result in potential theory
(uniqueness of solutions to the Dirichlet problem): There exists a
unique Green's function $G(\vec{r}, \vec{r}_0 )$, defined for $\vec{r}$,
$\vec{r}_0$ outside $B_{R_1}$, such that
$G$ satisfies $\nabla^2 G(\vec{r},
\vec{r}_0 ) = \delta(\vec{r} - \vec{r}_0 )$ and vanishes for
$\vec{r} \in S_{R_1}$ and for $\vec{r} \in S_{\infty}$
[as discussed, e.g., in \cite{Jackson},
for the two-sphere $S_{R_1}$ $G(\vec{r}, \vec{r}_0 )$ can be
constructed explicitly using the classic ``method of images"].
Plugging such a $G(\vec{r}, \vec{r}_0)$ into Eq.\,(4) as $V$ and
taking $\Phi(\vec{r})$ as $U$ and the region $B$ as the region {\em
outside} $B_{R_1}$ we obtain, by virtue of the vanishing boundary conditions
at infinity,
\begin{equation}
\Phi(\vec{r}_0 ) = \int_{S_{R_1}} \Phi(\vec{r}) \frac{\partial G(\vec{r},
\vec{r}_0)}{\partial n} \, d \sigma \; .
\end{equation}
Therefore $\Phi$ everywhere outside $B_{R_1}$ is determined uniquely by its
values on the two-sphere $S_{R_1}$.

We can now understand the kernel Eq.\,(2) in
the following way: Since the data for $\Phi$ consist of the values of a function
defined on a two-surface $S_{R_1}$, we can infer from these data uniquely
at the most another function of two variables, and not the full
three-dimensional density field $\rho(\vec{r})$. In fact, the forward kernel (or
the ambiguity in the corresponding inversion)
as described by Eq.\,(2) corresponds precisely to this geometric statement.

~

{\bf \noindent Weighing the shape of a body of known radial density}

One might hope that practical (physical) prior constraints on
the three-dimensional density distribution $\rho(\vec{r})$ might make it
uniquely recoverable from its far-zone gravity field despite the fundamental
non-uniqueness of the inverse problem. For example, we want the density
to be positive everywhere, which is a requirement that constrains the
ambiguity Eq.\,(2) to some extent. However, simple spherically-symmetric
counterexamples show that positivity is not a sufficiently strong
constraint to help provide us with a unique inversion. As the next step in
a series of physically-reasonable constraints on $\rho$, we might assume
a known positive radial density distribution
with profile $\rho(\vec{r}) \equiv \rho_0 (r) > 0$ distributed on
some arbitrary compact three-dimensional region $D$ in $\BR^3$. Put another way,
such a density profile represents
a body of arbitrary shape carved out of a spherically symmetric
(hence spherical) mass distribution. Again,
counterexamples based on hollow spherical shells show that this is not
quite enough for unique inversion. Nevertheless, it turns out that if we further constrain
the region $D$ such that it is connected and has no ``holes" (i.e., if $D$ is
topologically a ball), and, furthermore, if $D$ is ``radially convex"
in a sense made precise below, then unique inversion is possible:

\renewcommand{\thefootnote}
{\ensuremath{\fnsymbol{footnote}}}
\addtocounter{footnote}{1}

{\noindent \bf \underline{Theorem 3}}: Let $D$ be compact region in $\BR^3$ such
that its boundary $\partial D$ is a connected and simply-connected
surface (in other words, $\partial D$ is a topological two-sphere) which
is {\em radially convex} in the sense that any straight line in $\BR^3$ passing
through the center-of-mass of
the volume $D$ intersects $\partial D$ at precisely two
points. Assume that $D$ is filled with material of a known non-negative mass
density $\rho(\vec{r})$ which, when it is nonzero, is distributed
spherically-symmetrically with respect to the coordinate
origin given by the center of mass. That is,
if $\vec{r}$ lies inside $D$, then
$\rho(\vec{r}) = \rho_0 (r) > 0$, and if $\vec{r}$ is outside $D$, then
$\rho(\vec{r}) = 0$. Under these conditions, $D$ itself (or, equivalently, its
boundary $\partial D$) is uniquely recoverable from the far-zone gravity
field of this radial density distribution.\footnote{The
assumption that $\partial D$ is a topological
two-sphere is redundant since it follows from the assumption of radial
convexity as formulated in the theorem. However, it is perhaps useful to
emphasize this assumption in a redundant statement since the theorem is
certainly false without it.}

This result is not too surprising in view of Theorem 2, since
the specification of $\partial D$ entails just a single real function on
the two-sphere $S^2$ (measuring just how much we need to deform $S^2$ in order
to stretch it onto $\partial D$). The forward map Eq.\,(1) can then be
interpreted as a nonlinear map from real functions on $S^2$
(representing the deformations of $S^2$ needed to obtain $\partial D$)
to real functions on $S^2$ (representing the values of the potential $\Phi$ on $S_R$),
and we will now show that this map is locally one-to-one.

\noindent {\bf Proof of Theorem 3:}
The main idea of the proof is simple: explicitly write down,
in terms of spherical-harmonic coefficients,
the forward transform mapping the ``shape function" of $\partial D$ to
the exterior potential $\Phi$, and
show that the derivative of this nonlinear forward map
is nonsingular. The result then follows from the
inverse function theorem as generalized to infinite-dimensional
spaces~\cite{Marsden}. In this paper we will give a detailed
proof that
the forward map has nonsingular derivative at the point (shape) which
corresponds to a perfect sphere, so the result holds for shapes which are
nearby distortions of a perfect sphere (in other words, we will
explicitly prove that the forward
map is invertible in some open neighborhood of the perfect sphere in the
space of all shapes $D$ which satisfy the conditions of
the theorem). This case covers most planetary bodies
at the levels of resolution we are interested in.
Nevertheless, the statement that the
forward map is nonsingular everywhere remains valid,
although we are not going to
give an explicit proof of it here. The proof of this more general
case is substantially similar apart from requiring more careful estimates.

To proceed with the proof, introduce a spherical coordinate
system $( r, \theta , \phi )$ centered at the center of mass of the volume
$D$. In this coordinate system, let
${\bf n}(\theta , \phi )$ denote the half-line which starts at the
origin and expands outward in the direction $(\theta , \phi )$.
Let $\psi (\theta , \phi )$ be the (positive) function which gives the length of
the radial vector which starts at the origin and ends
at the intersection point of the line ${\bf n}(\theta , \phi )$
with the boundary $\partial D$ as $(\theta , \phi )$ ranges over
the unit two-sphere $S^2$ of all possible directions [by the radial
convexity assumption, there exists a unique such intersection point for
each direction $(\theta , \phi )$]. The function $\psi (\theta , \phi )$
can then be taken to be the ``shape function" which specifies $D$, and,
explicitly, we can write
\begin{equation}
D = \{ (r, \Omega) \; | \; r \leq \psi ( \Omega ) \} \; ,
\end{equation}
and
\begin{equation}
\partial D = \{ (r, \Omega ) \; | \; r = \psi ( \Omega ) \} \; ,
\end{equation}
where we introduced the short-hand notation $\Omega \equiv (\theta ,
\phi )$ for the angular coordinates. The exterior gravitational potential
$\Phi$ can be expanded in spherical harmonics\cite{Jackson}:
\begin{equation}
\Phi(\vec{r}) = \sum_{l,m} d_{lm} \frac{Y_{lm} (\Omega )}{r^{l+1}} \;
\; \; \; \; \; \; \; {\rm for} \; r>R> \max_{\Omega} \psi(\Omega ) \; ,
\end{equation}
where we can regard $\{ d_{lm} \} \equiv {\bf D}$ as an infinite sequence
(vector) of ``observables" which completely describes the data for
the inverse problem in view of Theorem 2. On the other hand, according to Eq.\,(1),
for $r>R> \max_{\Omega} \psi(\Omega )$ we have
\begin{eqnarray}
\Phi(\vec{r}) & = & G  \int_{r' < \psi(\Omega ' )} \rho_0(r') \, \frac{d^3 r'}
{| \vec{r} - \vec{r}\, ' |} \nonumber \\
& = & G \int_{0}^{\psi( \Omega ')} \rho_0(r') \, r'^2 \, dr' \int_{S^2}
\sum_{l,m} \frac{r'^l}{r^{l+1}} Y_{lm}(\Omega) Y^{\ast}_{lm}(\Omega ')
\, d \Omega ' \nonumber \\
& = & G 
\sum_{l,m} \frac{Y_{lm}(\Omega)}{r^{l+1}} \int_{S^2} Y^{\ast}_{lm}(\Omega ')
\; \mu_{l+2} \left( \psi(\Omega ') \right)
\, d \Omega ' \nonumber \\
& = & G  \sum_{l,m} f_{lm}[\psi ] \frac{Y_{lm} (\Omega )}{r^{l+1}} \; ,
\end{eqnarray}
where
\[
\mu_n (w) \equiv \int_{0}^{w} \rho_0 (r) \, r^n \, dr \; , \tag{10a}
\]
and
\begin{equation}
f_{lm} [\psi ] \equiv  \int_{S^2} 
Y^{\ast}_{lm}(\Omega ) \; \mu_{l+2} \left( \psi(\Omega ) \right) \, d \Omega \; 
\end{equation}
is a vector functional of the shape function $\psi$ which represents the
forward map in the same way as ${\bf D} = \{ d_{lm} \}$ represents the
data. In fact, introducing the notation
${\bf F} [\psi ] \equiv \{ f_{lm} [ \psi ] \}$ and combining Eqs.\,(8)
and (9), the forward equation for the shape function $\psi$ takes the simple form
\begin{equation}
{\bf F} [\psi ] = \frac{1}{G} \; {\bf D} \; .
\end{equation}
(Due to our choice of origin as the center of mass, both $f_{lm}$ and
$d_{lm}$ vanish for $l=1$, but this fact will not be of any consequence in
what follows.)  It is also convenient to
introduce a coordinatization of the space of shape functions via a
spherical harmonic expansion
\begin{equation}
\psi(\Omega ) \equiv \sum_{l,m} s_{lm} Y_{lm}(\Omega ) \; ,
\end{equation}
and consider the coordinate vector ${\bf S} \equiv \{ s_{lm} \}$ as the
representation of the function $\psi (\Omega )$. In this coordinate
system the forward map Eq.\,(11) takes the form
\begin{equation}
{\bf F} [{\bf S} ] = \frac{1}{G} \; {\bf D} \; ,
\end{equation}
where
\begin{equation}
f_{lm} [{\bf S}] \equiv  \int_{S^2} 
Y^{\ast}_{lm}(\Omega ) \; \mu_{l+2} \left(
\sum_{p,q} s_{pq} Y_{pq}(\Omega ) \right) d \Omega \; .
\end{equation}
Assume now, contrary to the conclusion of Theorem 3, that
two distinct domains $D_1$ and $D_2$ constrained as in the statement
of the theorem give rise to identical external
gravitational potentials when filled with the
given radial density distribution $\rho_0 (r)$.
First of all, since the monopole and dipole moments
of the two mass distributions must agree, they must have
the same center of mass, therefore we can set up a common spherical
coordinate system for both volumes with their shared center of mass
chosen as the origin of coordinates. It then follows
that there exist two distinct shape functions $\psi_1$ and $\psi_2$, corresponding
to the two distinct volumes $D_1$ and $D_2$, which
satisfy Eq.\,(11) with the same data $\bf D$; in other words
\begin{equation}
{\bf F} [\psi_2 ] = {\bf F} [\psi_2 ] \; .
\end{equation}
We will now show that Eq.\,(15) is impossible as long as $\psi_1$ and
$\psi_2$ belong to some fixed open neighborhood
of a perfect sphere $\{ \psi (\Omega ) \equiv a_0
= {\rm const} \}$ in the infinite-dimensional nonlinear function space
of all $\psi$'s. Using the inverse function theorem as generalized to
such infinite-dimensional manifolds~\cite{Marsden}, it is sufficient to
show that the derivative of the map $\bf F$ at the point $\psi(\Omega )
\equiv a_0 $ is a nonsingular linear map. In general,
at an arbitrary point $\psi = \psi_0 (\Omega)$, this derivative
is given by
\begin{equation}
({\bf F}' [\psi_0(\Omega)] \cdot \delta \psi)_{lm} = \int_{S^2} 
Y^{\ast}_{lm}(\Omega ) \; \rho_0 \left(
\psi_0 (\Omega ) \right) \;
\psi_0 (\Omega )^{l+2} \; \delta \psi
(\Omega ) \, d \Omega \; ,
\end{equation}
where ${\bf F}' [\psi_0(\Omega)]$ denotes the derivative evaluated at
the point $\psi = \psi_0$,
acting (as a linear map) on the tangent vector (linear perturbation)
$\delta \psi$, and we have used Eq.\,(10) to derive this explicit form.
Specializing to the perfect sphere $\psi_0(\Omega ) = a_0 (= {\rm const})$
and using the coordinate representation [cf.\ Eq.\,(12)]
\begin{equation}
\delta \psi(\Omega ) \equiv \sum_{p,q} \delta s_{pq} Y_{pq}(\Omega ) \; ,
\; \; \; \; \; \; \delta {\bf S} \equiv \{ \delta s_{pq} \} \; ,
\end{equation}
Eq.\,(16) takes the form
\begin{eqnarray}
({\bf F}' [a_0] \cdot \delta {\bf S})_{lm} & = & \rho_0 (a_0 ) \, {a_0}^{l+2} \int_{S^2} 
Y^{\ast}_{lm}(\Omega ) \left[ \sum_{p,q} \delta s_{pq} Y_{pq}(\Omega )
\right] \, d \Omega \; \nonumber \\
& = & \rho_0 (a_0 ) \, {a_0}^{l+2} \delta s_{lm} \; ,
\end{eqnarray}
where we made use of the fact that the $Y_{lm}$'s form an orthonormal
basis for $L^2 (S^2)$. According to Eq.\,(18), the derivative ${\bf F}' [a_0]$
is a {\em diagonal} linear map with only nonzero entries
(eigenvalues) on the diagonal; therefore, ${\bf F}' [a_0]$
is clearly nonsingular. This completes the proof of Theorem 3.

~

{\bf \noindent The kernel of the forward map onto gravity gradient observables}

The gravitational gradient tensor is (apart from a minus sign)
simply the (symmetric) tensor of
second derivatives of the potential $\Phi$
in a cartesian coordinate system:
\begin{equation}
T_{ij} \equiv - \frac{\partial^2 \Phi}{\partial x^i \partial x^j} \; .
\end{equation}
So, for example, we have
\begin{equation}
T_{xx} = - \frac{\partial^2 \Phi}{\partial x^2} \; ,
\; \; \; \; \; \; \; T_{yz} = -
\frac{\partial^2 \Phi}{\partial y \, \partial z} \;
\end{equation}
etc. Independently of coordinates, the gradient tensor
can be defined as the double covariant derivative
$\nabla \nabla \Phi$ (in general relativity, $T_{ij}$ corresponds to the
Riemann curvature tensor $R_{0i0j}$ describing
tidal gravitational forces).
One can also define the gradient tensor explicitly in terms of the
source mass distribution as:
\begin{equation}
T_{ij} (\vec{x}) = G \int \frac{\rho(\vec{y}) \,
[\, 3 \, (x^i - y^i ) (x^j - y^j ) - \delta_{ij} \, |\vec{x} - \vec{y}|^2 \, ]}
{|\vec{x} - \vec{y}|^5}
\; d^3 y \; ,
\end{equation}
where all coordinates are cartesian. The gradient tensor is
a particularly useful observable in precision gravimetry since
it is better isolated from local non-gravitational acceleration noise
compared to other observables,
and a large roster of instruments (gradiometers) are available for measuring it.

In practical applications, one often works
with a coordinate system where $z$ is the vertical
coordinate pointing up from the Earth's center, and the observable
of interest is the $x$---$y$ projection of the gradient tensor in an infinitesimally
small neighborhood (tangent plane to the Earth's spherical surface) around $x=y=0$:
\begin{equation}
T \equiv 
\begin{pmatrix}
   T_{xx} & T_{xy} \cr
   T_{xy} & T_{yy}
  \end{pmatrix}
\end{equation}
Typically, a gradiometer takes two kinds of measurements: 
the component $M_{\times} \equiv 2 T_{xy}$ (``crossline"),
and the combination $M_{+} \equiv T_{xx} - T_{yy}$ (inline).
The choice of cartesian $x,y$ coordinates is arbitrary upto a rotation $R$, and
$T$ transforms under rotations as
\begin{equation}
T \longrightarrow R \, T R^{t} \; .
\end{equation}
Neither $T$ nor its crossline
or inline components are invariant under rotations, but
$\rm{Tr} (T)$ and $\rm{Det} (T)$ are invariants.
In particular, the Euclidean norm of $(M_{+}, M_{\times})$
\begin{equation}
\sqrt{M_{+}^2+M_{\times}^2} = \sqrt{ \mbox{Tr} (T)^2 - 4 \, \mbox{Det} (T) } \; 
\end{equation}
is an invariant. More specifically, it is easy to compute that
$(M_{+}, M_{\times})$ transforms under a rotation
\begin{equation}
R_{\theta} = 
\begin{pmatrix}
   \cos \theta  & - \sin  \theta  \cr
   \sin  \theta  & \cos  \theta 
  \end{pmatrix}
\end{equation}
according to the rule
\begin{equation}
\begin{pmatrix}
   M_{+}  \cr
   M_{\times}  
  \end{pmatrix}
  \longrightarrow R_{2 \theta}
  \begin{pmatrix}
   M_{+}  \cr
   M_{\times}  
  \end{pmatrix} \; .
\end{equation}
Now, define a new observable
\begin{equation}
V \equiv M_{+}^2+M_{\times}^2 \; .
\end{equation}
In other words, $V$ is defined at every point of 
${\BR}^3$ by setting up local coordinates $x,y,z$ such that the $z$-axis goes
through the origin, computing the gradient tensor $T$ as in Eqs.\,(19) and (22) at
that point, and then calculating the norm square of the
observable $(M_{+}, M_{\times})$. The key features of $V$ are that
(i) it is invariant under rotations,
therefore uniquely defined independently of the choice of coordinates, and
(ii) it obeys the following lemma:

\noindent {\bf Lemma:} $V$ vanishes at a point if and only if  $(M_{+}, M_{\times})$
vanishes there for any allowed choice of coordinates $x,y,z$.

\noindent {\bf Proof:} By Eq.\,(11), $V$ is invariant and equal to
the norm-square of $(M_{+}, M_{\times})$ for any choice of coordinates.

Therefore the kernel of the coordinate-dependent
observable $(M_{+}, M_{\times})$ is precisely the
kernel of the nonlinear but coordinate-independent observable $V$.

\noindent {\bf \underline{Theorem 4:}} Let $F[\Phi ]$ be any analytic functional (linear
or nonlinear) of the
gravity potential $\Phi $ (such as $V$), and let $S$ be any analytic 2-surface lying
outside the spherical region $B_R=\{\vec{r} : |\vec{r}| < R \}$ in ${\BR}^3$
(such as a sphere of radius $> R$). Then, if $F[\Phi ]$ vanishes in any open
(two-dimensional) neighborhood on $S$, it vanishes identicallly on all of $S$.

\noindent {\bf Proof:} This just follows from analyticity: $\Phi$ is an analytic
function outside the spherical region $B_R=\{\vec{r} : |\vec{r}| < R \}$ in ${\BR}^3$
since it satisfies the homogeneous Laplace equation there (analyticity follows
from standard elliptic regularity theorems~\cite{ellipreg}). Therefore,
$F[\Phi ]$ is analytic there and so is its restriction to $S$ since $S$ is
analytic. Thus vanishing on any open subset is equivalent to vanishing
identically on $S$.

\noindent {\bf Corollary:} If $V$ vanishes in
any two-dimensional patch, no matter how small, on any analytic
observation surface $S$ lying
outside the spherical region $B_R=\{\vec{r} : |\vec{r}| < R \}$,
then it vanishes identicallly on all of $S$.

This corollary further illustrates the fact that the kernel of
the gravity-gradient observable $(M_{+}, M_{\times})$ is precisely the
kernel of the observable $V$, not only locally but also globally.

\noindent {\bf \underline{Theorem 5:}} Let $S$ be a sphere of radius $> R$. Then $V$ vanishes
on $S$ if and only if $\Phi$ is spherically symmetric
(a function of the radius $r$ only), and hence
$V$ vanishes identically everywhere outside the spherical region $B_R=\{\vec{r} : |\vec{r}| < R \}$
in ${\BR}^3$.

\noindent {\bf Proof:} The if part is a simple calculation:
It is straightforward to compute that both
$M_{+}$ and $M_{\times}$ vanish for a radial (monopole)
potential function $\Phi(r)$. For the converse, it is easy
to see that if $V$ vanishes on a sphere
$S$ then this implies that $\Phi$ is constant on $S$. But this implies,
according to Eq.\,(5) (or just by the uniqueness of solutions to the Dirichlet
problem), that $\Phi$ is a radial function of monopole type:
\begin{equation}
\Phi (\vec{r} ) = \frac{C}{r} \; ,
\end{equation}
where $C$ is a constant.
 
We can now completely characterize the kernel of the forward
map from the mass density to the gravity gradient observables
$(M_{+}, M_{\times})$:

{\noindent \bf \underline{Theorem 6}}: The kernel of the forward map
mapping mass distributions $\rho$ supported in $B_R$ to gravity gradient observables
$(M_{+}, M_{\times})$ outside the region $B_R$ (i.e.\
for $|\vec{r}|>R$)
is precisely functions $\rho$ supported in $B_R$ satisfying
\begin{equation}
\rho = \rho_0 \, + \, {\nabla}^2 \chi \; ,
\end{equation}
where $\chi(\vec{r})$ is any (sufficiently
smooth) function on $\BR^3$, and $\rho_0$ is any {\em spherically
symmetric} function, both supported inside $B_R$ (i.e., both
$\chi(\vec{r})$ and $\rho_0 (\vec{r})$ vanish for $r>R$).

\noindent {\bf Proof:} By Theorem 5, the kernel of the
map from mass distributions to outside gradients $(M_{+}, M_{\times})$
consists of those mass distributions that give rise to spherically
symmetric potentials $\Phi$ outside $B_R$.
If $\Phi$ is spherically symmetric outside $B_R$,
then consider $\Phi_S$, the spherical
average of $\Phi$ (i.e.\ $\Phi_S (\vec{x})$ = the average
of $\Phi$ on the sphere centered at 0 and passing through $\vec{x}$).
Then $\Phi_S$ is spherically symmetric everywhere and coincides
with $\Phi$ outside $B_R$. Therefore $\chi \equiv \Phi - \Phi_S$ vanishes
outside $B_R$, and thus
\begin{equation}
\nabla^2 \Phi = \nabla^2 \chi + \nabla^2 \Phi_S \; .
\end{equation}
Since $\Phi_S$ is everywhere spherically symmetric, so is
$\nabla^2 \Phi_S$, and Theorem 6 follows.

~


{\bf \noindent Kernel of the gravitational forward map and discretization}

In practice, gravity inversion is a discrete problem because (i) the measurements
of $\Phi$ (or of the gradients)
are finitely many and discretely distributed in space, and (ii)
more problematically, the model for the mass distribution $\rho$ is some
discretized approximation to a continuous distribution. In all
contexts, the characterization of $\rho$ would be a finite list of
parameters which uniquely specify $\rho$
in some generally non-linear fashion. For example, these $\rho$ parameters could
be the masses, locations, and shape parameters
of a finite number of tectonic plates in a geophysical model of the
Earth's crust. Or, as discussed above, they could be masses of finite
blocks into which we divide the source distribution in discretizing it.
Or, more straightforwardly, they could be the masses of $N$
point-mass centers distributed throughout the source region,
approximating with a discrete configuration the true mass distribution
in the limit $N \rightarrow \infty$.

In general, the practical, the discretized gravitational
inverse problem is the problem of inverting
some (generally nonlinear) forward map:
\begin{equation}
F: \; \{ p_j \} \longmapsto \{ \Phi_i \} \; , \;
\; \; \; \; \; \Phi_i = F_i [p_j ] \; ,
\end{equation}
where $p_j$ are finitely many parameters specifying the mass distribution,
and $\Phi_i$ are the measurements. It would be conceptually salutary
to have the fundamental non-uniqueness in the gravitational
inverse problem (the kernel of the froward map) described by
Theorem 1 to fall out of the formulation Eq.\,(31
) in a natural way. For
example, when we simulate a slab of soil using some large number of mass
centers regularly placed at fixed lattice points inside the slab, the
parameters $p_j$ are simply the point masses $m_j$ assigned to each center at
lattice location $j$. In this case, the forward map $F$ is in fact linear:
\begin{equation}
\Phi_i = \sum_j F_{ij} m_j \; ,
\end{equation}
where the matrix $F_{ij}$ is the Green's function in Eq.\,(1) in
discretized form:
\begin{equation}
F_{ij} = - G \frac{1}{|\vec{r_i} - \vec{r_j}|} \; ,
\end{equation}
with $\vec{r_i}$ being the locations where the measurements $\Phi_i
\equiv \Phi (\vec{r_i})$ are collected. Consider first, for simplicity, a
scenario in which we are sampling $\Phi$ at the same number of points
$N$ as the number of mass centers in the discretization. In other words,
$F$ is now a square $N \times N$ matrix. In view of Theorem 1 characterizing
the kernel of the forward map,
one might expect $F$ to be singular, with the null space corresponding to a
discretized version of the kernel, i.e., a discrete approximation to
functions of the form $\nabla^2 \chi$
with $\chi$ supported inside the slab. It turns out, however, that
the matrix $F$ given by Eq.\,(33) is in fact generically nonsingular.
Moreover, the $M \times N$ matrix $F$ with $M$ measurement locations and
$N$ mass centers is also nonsingular, in the sense that generically it
has maximal rank (i.e.\ trivial null space).

It is in fact easy to see why this is so, because of the following result:

{\noindent \underline{\bf Theorem 7}}: Given
a solution $\Phi(\vec{r})$
of the Laplace equation ${\nabla}^2 \Phi=0$ vanishing at infinity
and defined for $r>R$, there exists {\it at most} one configuration
$\{ m_j , \vec{r_j} \}$ of
finitely many point masses placed inside $B_R = \{ r \leq R \}$ (i.e.\
with $m_j \in \BR$ and $| \vec{r_j} | \leq R $) that can give rise to
this $\Phi$ for $r>R$.

{\noindent \bf Proof}: Suppose, on the contrary, that there are two
configurations of point masses,
$\{ m_j , \vec{r_j} \}$ and $\{ m_k ' , \vec{r_k}' \}$,
that produce the same $\Phi$ for $r>R$. Subtract the second
configuration from the first, and correspondingly subtract
the $\Phi$ fields that they produce. Since the gravity field depends
linearly on the mass distribution, what we obtain is a new configuration
$\{ m_1, \cdots , m_N, -m_1 ' , \cdots -m_{N'} ' , \vec{r_1},
\cdots , \vec{r_N}, \vec{r_1}' , \cdots , \vec{r_{N'}} ' \}$ of point
masses inside $B_R$ (unless there are some coincident point masses in
the two collections, in which case one would simply subtract the
corresponding masses
and list the location only once), which produces a field $\Phi$ that
vanishes identically for $r>R$. Could this actually happen? It turns out
the answer is no, unless $\Phi$ is identically zero everywhere (and
therefore the two original point-mass configurations are in fact
identical). To see this, observe that $\Phi$ produced by a finite set of
point masses is a real-analytic function in $\BR^3$ except at the locations
of the point masses where it has singularities. Since $\Phi$ vanishes
for $r>R$ and is analytic,
it must vanish everywhere in $\BR^3$ except possibly at the
mass centers. But if any of the mass centers had non-zero mass, we could
choose points so close to that center that the contribution to
$\Phi$ from that center would overwhelm the contributions from any other
centers (which are discretely spaced since there are finitely many).
This clearly contradicts the fact that $\Phi$ is identically zero in any
small neighborhood of the chosen mass-center. Therefore, none of the mass
centers can have nonzero mass; the two original configurations of
point masses must be identical, and Theorem 7 is proved.

Here is one way to understand the apparent conflict between
Theorem 1 and Theorem 7: Consider the two spaces between which the forward map
$F$ acts: the space of density distributions $\rho$
and the space of potentials $\Phi$. Any discretization is an attempt to
approximate these spaces via a sequence of finite-dimensional subspaces. For
example, when we use $N$ point masses, we have an $N$-dimensional subspace
of the space of all $\rho$, and as $N$ gets larger and larger this subspace
approximates the full space arbitrarily closely, in the sense that for
any $\rho_0$, we can find a configuration of $N$ point masses (with large
enough $N$) which comes as close as we want to $\rho_0$ (in some locally
averaged sense). The same goes for
the corresponding potentials $\Phi$: given any solution $\Phi_0$, we can
find potentials produced by $N$ point masses that get arbitrarily close
to $\Phi_0$ as $N \rightarrow \infty$. But the problem is that these
approximating subspaces completely miss the kernel of the true forward
map, which is the subspace of mass distributions (and corresponding
potentials) given by $\{ \nabla^2 \chi \; | \; \chi \in C_0(B_R ) \}$.
The intersection of the approximating subspaces with this kernel
subspace is the zero vector, for any finite $N$. This is
(mathematically) the explanation for the apparent contradiction
between Theorem 1 and Theorem 7.

To resolve this apparent conceptual paradox,
one might argue that we must choose the approximating
finite dimensional subspaces in such a way that they fully intersect the kernel.
But in practice, there is no feasible
way to discretize the problem that
makes sure this property holds. There is, however,
a much simpler practical strategy out of this apparent paradox,
and this is the strategy we advocate: Realize that true
measurements in the real world always have instrumental noise. What this means is
that two potentials are indistinguishable in practice if they differ
everywhere by less than, say 1$\sigma$ worth (in some
arbitrary units) of instrumental noise.
Therefore, e.g.\ when we look for the intersection between the kernel
and our discretized
$\rho$-subspace with $N$ point masses, what
we are really looking for are all $N$-point-mass
configurations that produce a potential $\Phi$ that differs from zero by
less than $1\sigma$ throughout the exterior region $r>R$. And in general
there are many such configurations. We can see this, for example, in the
matrix $F_{ij}$ of Eq.\,(33): in general this matrix turns out to be highly
ill-conditioned (with very small determinant) with lots of eigenvalues close
to zero, even though it has no exactly-zero eigenvalues. And the
``approximately null" subspace spanned by the small-eigenvalued eigenspaces
is precisely the discrete analogue of the kernel of the
forward map; it is what corresponds to the subspace
$\{ \nabla^2 \chi \; | \; \chi \in C_0(B_R ) \}$ in this discretization.
We expect a similar description for the discrete analogue of the forward map's kernel
in any other practical discretization scenario.

%
%

%
%
%

\end{document}